\documentstyle{article}
\begin{document}
\begin{titlepage}
\centering {\Large\bf  INTERNATIONAL CENTER FOR THEORITICAL
PHYSICS \\ }
\vskip 3truecm {\large \bf  Cohomological Quantum Mechanics\\
And\\ Calculability of Observables}\\
\vskip 1truecm {\large \bf  M.MEKHFI\\}
\vskip 8truecm {\bf Trieste ITALY\\} December 1995
\end{titlepage}
\title{\bf Cohomological Quantum Mechanics\\ And \\
Calculability of Observables }
\author { {\bf  M.MEKHFI} \thanks {On sabbatical leave from
Institut de Physique Univ Es-Senia Oran-ALGERIE .Work supported
by ICTP }\\
\normalsize International Center For Theoritical
Physics.Trieste, ITALY}
\date {23 December 1995}
\maketitle
\vskip 4truecm
\begin{abstract} We reconsider quantum mechanical systems based
on the classical action being the period of a  one form over a
cycle and elucidate three main points.First we show that the
prepotenial V is  no  longer completely arbitrary but obeys a
consistency integral equation.That is the one form dV  defines
the same period as the classical action.We then apply this to
the case of the punctured  plane for which the prepotential is
of the form $ V= \alpha \theta + \Phi ( \theta ) $ .The function 
$ \Phi $ is any but a periodic function of the polar angle .For
the topological information to be  preserved , we further
require that $ \Phi $ be even.Second  we point out the existence
of a  hidden scale which comes from the regularization  of the
infrared behaviour of the solutions.This  will then be used to
eliminate certain invariants preselected on dimensional
counting  grounds.Then provided we discard nonperiodic solutions
as being non physical we compute the  expectation values of the
BRST- exact observables with the general form of the
prepotential  using only the orthonormality of the solutions (
periodic ).Third  we give  topological  interpretations of the
invariants in terms of the topological invariants wich  live 
naturally on the  punctured plane as the winding number and the
fundamental  group of homotopy,but this  requires a prior
twisting of the homotopy structure
\end{abstract}
\newpage
\section {Introduction} The interest on studying cohomological
quantum mechanical systems  here is to  understand as a first
step ,the way the supersymmetry and the ghost number are
broken        \cite{kn:Baulieu} in topological field theories,
and the way to compute expectation values of the inherent
topological  invariants.In the class of topological field
theories which can be rephrased as gauge fixed  topological
actions \cite {kn:Witten}  \cite{kn:Baulieu}
\cite{kn:Thompson},observables are shown to be BRST- exact and
have non vanishing  expectation values ,but that presupposes
that the underlying symmetry is somehow broken in a 
 special way .To investigate these systems from the above point
of view several difficulties   arise.First the gauge fixing
procedure through the use of a priori arbitrary prepotential V
may  spread out the initial topological information contained in
the classical action.Second a hidden  scale is shown to enter
the analysis through the infrared regularization of the
wavefunctions and this  makes dimensional counting necessary but
rather insufficent to preselect topological  invariants.Third
the identification of the observables with natural topological
invariants needs first a  kind of twisting of the homology
structure.In this paper we investigate and elucidate these
points.The starting point is   the cancellation  of the
classical action during the process of the gauge fixing .This is
a necessary requirement for the gauge fixed action to possess a
secondary   (dual) BRST symmetry,otherwise one cannot rewrite
the hamiltonian in the form H $ \propto \{ Q,\bar Q \}  $with 
Q  and $  \bar Q $ both nilpotent  .The prepotential is then
required  to belong to the class of functions such that the one
form dV defines the  same period as the classical action in
order to properly perform the above cancellation.In other words
we simply require that  for two  quantities to identically
cancel each other they should first be of the same
(cohomological)  nature.We then apply the idea to the case of
the punctured plane  and show that with the above selection
criteria there is no loss or spread of the topological 
information as we could define things unambiguously.To
eventually capture the right topological information  without
spurious backgrounds,the nonlinear term of the  the prepotential
in addition to its being periodic is further  required to be an
even function of the argument.The evenness of the prepotential
is an essential  input for the proper identification of the
observables and will be justified both topologically and
analytically in subsequent sections.On the other hand the
problem is highly non  trivial as it involves general
Sturm-Liouville systems.We will show however that as long as we
discard nonperiodic solutions which are seen not to reproduce
orthonormality ,full knowledge of the spectrum and the
associated eigenfunctions is hopefully not needed.The
orthonormality of the periodic solutions only   does really
matter as a consequence of the even parity of the potential.
The  hidden scale is shown to come from  Bessel functions which
are known to be normalized  only on finite "volume" ,hence a
lenth L is introduced .This new scale will subsequently be used 
to discard certain "topological" invariants preselected on
dimensional counting grounds.Finally the identification of the
invariants in terms of the winding number W and the elements of
the  fundamental group of homotopy $\Pi ^ s( m ) $  of the
punctured plane calls for a type of twisting of the homotopy 
structure.The reason one has to do that is that the invariants
to be identified  being  Q-dependant (Q-exact ) depend
explicitly on  the coupling constants present in the theory as
Q$ \propto \partial V$  but these extra parameters ( coupling
constants ) are not needed to characterize the (algebraic)
topology of the punctured plane so one has to introduce them
into the homotopy structure in one way or another.In an
unpublished work \cite{kn:Mekhfi} we performed such twisting of
the homotopy structure  by introducing an appropriate self
interaction between homotopic loops viewed momentarily as 
intrinsec physical objects which may possibly  self interact.In
the process of identification and in  the case of the two  cases
we studied we arrived at the following identifications of the 
invariants $ W+i\alpha 1$ and $ J+i\alpha 1 $ with $ J=- \sum
_{m\epsilon Z/0} \rho (m)  \frac{\Pi (m)}{m} $ where W is the 
winding number ,1 is the identity of the group of homotopy and
the $\Pi^s ( m ) \; m \neq 0 $ are the rest of the  group
elements.$\rho (m)$is an m-dependant coupling constant with had
acquired a priori a  meaning in the context of a twisted 
homotopy.
\section{Gauge fixing topological actions} Let M be a compact
manifold with local coordinates $ x^{\mu} $ and let $ H^r ( M )$
be the $r^{th} $ de Rham cohomology group .If $ c_1,......c_k $
are elements of the homology group $ H_r( M ) $ with k the $
r^{th} $ Betti number such that $ [c_i] \neq [c_j] $, then for
any set of numbers $b_1,...b_k $  ,a corrollary of the de Rham's
theorem states  that there exist a closed r-form  $ \omega$ such
that.
\begin {equation}
\int_{c_i} \omega = b_i \;\;\;\;\; 1\leq i  \leq k
\label {eq:action}
\end {equation}
\noindent We may include in this expression the case $ b_i=0$
for which $ \omega$ is closed and exact  .The numbers in
~\ref{eq:action} are the periods of closed r-forms over cycles
$c_i$.Cohomological quantum  mechanics is defined as a system
whose classical action is the period $b_1$ .The main feature  of
such an action is that it is defined on $ H_r(M)\;X\;H^r(M) $and
is therefore topological,that is  invariant under any
infinitisimal deformation $ \epsilon $ which keeps the cycle
within its  homology class.
\begin {equation}
\delta x^\mu=\epsilon^\mu
\label {eq:deformation}
\end {equation}
\noindent The BRST symmetry associated to the above symmetry can
be chosen as simple as.
\begin {eqnarray}
			    sx^\mu &=&\psi^\mu \nonumber \\
			  s\psi^\mu& =&0 \nonumber \\
		     s\bar{\psi^\mu}&=&\lambda ^\mu \\
		   s \lambda ^\mu&=&0 \nonumber 
\label {eq:BRST}
\end {eqnarray}
\noindent Provided one chooses a quite involved gauge
function.To gauge fix the symmetry such that  covariance is
maintained and to get an action quadratic in velocities Baulieu
and Singer \cite {kn:Baulieu1} proposed the following gauge
function.
\begin {equation}
\dot{x^\mu}+\frac{\partial V}{\partial x^\mu} +...
\label {eq:Gauge function}
\end {equation}
\noindent Where $ \dot x^\mu =\frac{dx^\mu}{dt} $ and where we
omit  the  Christoffel symbol term as  these are known matters
and not relevant for what follows.We will  concentrate in this
section on the prepotential  V .The  prepotential is a priori an
arbitrary given function of the coordinates $ x^\mu$ .The
expectation  values of the invariants may depend on the form of
the class of the functions V as the Euclidean  path integral
probes the moduli space of the equation $ 
\dot{x^\mu}+\frac{\partial V}{\partial x^\mu}$ .In this paper we
want to show among other things that the prepotential is not
that arbitrary but is restricted by a consistency equation.Write
the gauge fixed action .
\begin {equation} S_{SG}= \int_{c\epsilon H_1} \omega +\int
dt\;\; s\; \bar {\psi}^\mu ( g_{\mu \nu}
\dot{x}^\nu+\frac{\partial V}{\partial x^\mu} - \frac{1}{2}
g_{\mu\nu} \lambda ^\nu+...) 
\label {eq:fixed action}
\end {equation}
\noindent After integrating out the auxillary field $\lambda$ we
select out the bosonic linear term of interest
\begin {equation}
\int \dot{x}^\mu \frac{ \partial V}{\partial x^\mu} dt =\int dV 
 \label {eq:prepotential1}
\end {equation}
\noindent The second step is to cancel out the classical
action.This cancellation is a necessary requirement  for the 
resulting gauge fixed action to possess a secondary ( dual )
BRST symmetry which in turn allows  the hamiltonian to have the
required form H $ \propto \{Q,\bar Q \} $ with  Q  and $ \bar Q
$ both nilpotent.Now comes our remark that  in order for two
quantities to identically cancel each other  they should first
be of the same (cohomological) nature.Therefore the prepotential
V should be  such that the one form in ~\ref {eq:prepotential1}
is a period and that this equates the classical action.
\begin {equation}
\int_{c\epsilon H_1} \omega =\int_{c\epsilon H_1} dV
 \label {eq:consistancy}
\end {equation}
\noindent In other words ,one should not just look for a
prepotential on the basis that the integral of it  numerically 
cancels out the classical action.We will check in the case of
the punctured plane which we will  study in details how
restrictive the selection criteria is and how indeed it
encompasses the entire  topological infomation of the classical
action.A completly arbitrary choice of the prepotential will 
eventually spread out the initial topological information.We
hereafter specialize to the case of  the punctured plane $ 
R^2/(o)$ and take it  as our target manifold,as it has simple
but not trivial topology.In this case we  have .
\begin {equation} H^1 ( R^2/( 0 ) ,R ) \cong R \\
 \label {eq:cohomology}
\end {equation}
\noindent The $ \omega 's $ are then one forms labelled by real
numbers and the cycles c are homotopic loops  encirling the
whole as $  H_1 \cong \Pi_1 \cong Z   $  where $ \Pi_1$ is the
first homotopy group or the fundamental group and Z is the set
of integers.The topological action associated with the punctured
plane is , $ \alpha  \epsilon R $ \cite{kn:Nakahara}
\begin {equation} S_{cl}=\int_c \alpha \;\;d\theta \\
 \label {eq:action2}
\end {equation}
\noindent Where we define polar coordinates as
$x^1+ix^2=re^{i\theta} $.The solution to the equation  ~\ref
{eq:consistancy} for the case of the punctured plane is.
\begin {equation} V=\alpha \theta + \Phi ( \theta )\\
 \label {eq:prepotential}
\end {equation}
\noindent The function $\Phi ( \theta )$ is any but periodic
function of $ \theta$ .This form is entirely due to the 
cohomological nature of the punctured  plane.The simplest case $
\Phi =0$ has been selected 
 by Baulieu and Rabinovici \cite {kn:Baulieu} on the basis of
local BRST.Let us open a parenthesis about the use of  local
BRST in the present context as the latter is considered the
important input of reference \cite {kn:Baulieu}. Local BRST
symmetry provided the authors with an arrow of selective
differential equations for the prepotential which we rewrite in
a form  more appropriate for the discussion.( To be compared to
our selection criteria ~\ref{eq:consistancy} )     
\begin {equation}
\frac{\partial}{\partial x_j} ( x_i \frac { \partial V}{\partial
x_i })=0
 \label {eq:local brst}
\end {equation}
\noindent As it is written this equation simply means that the
prepotential $ V ( r,\theta ) $ is in fact   function of the
$\theta $ variable only and should not depend on the radial
variable r as one may write 
$ x_i \frac{ \partial V }{\partial x_i}=r \frac{\partial
V}{\partial r}  $. We therefore conclude that local BRST
invariance is poorely restritive as any  function of $\theta $
may well do and so it is not enough to select the particular
value $\alpha \;
\theta $ of the prepotential as the above authors did. In
subsequent sections we will solve the problem completely using
the full expression of the prepotential ~\ref {eq:prepotential}
including the $\Phi$ term.
\section {Bessel functions and the definition of observables} In
the following we will compute observables in the canonical
formalism.The hamiltonian  associated with the action ~\ref
{eq:fixed action},which we adapt to the punctured plane is
.\cite {kn:Baulieu}
\begin {eqnarray} H&=&\frac{1}{2} p^2 +\frac{1}{2r^2}
{(\frac{\partial V}{\partial \theta })}^2  -\frac{1}{2} [ \bar
{\psi_i},\psi_j ] \frac{ \partial^2 V}{\partial x_i \partial
x_j}      \nonumber \\
  &=&\frac{1}{2}  \{ Q,\bar Q \}  
\label  {eq:hamiltonian}
\end{eqnarray}
\noindent Where $  p_i=-i\frac{\partial}{\partial x_i}  $ and $
\bar {\psi_i} = \frac{\partial}{\partial \psi_i} $   are
canonical momenta for the coordinates $x_i$ and the ghost 
fields$\;\;\psi_i$  respectively and where the generators Q  and
$ \bar Q $ are given by.
\begin {eqnarray}
       Q &=&\psi_i ( p_i+i\frac{ \partial V}{\partial x_i}
)\nonumber \\
\bar Q &=&\bar {\psi_i}( p_i-i\frac{ \partial V}{\partial x_i} )
\label  {eq: generators}
\end{eqnarray}
\noindent Let us first consider the eigenvalue problem
associated to the hamiltonian ~\ref {eq:hamiltonian} without
specifying for the moment the form of the prepotential.For this
purpose  it is appropriate to adopt the superwavefunction
formalism.Let $\Phi (x,\psi)$ denotes the superwavefunction.
\begin {equation}
\Phi ( x,\psi ) = \phi + \psi_iA^i+ \frac{1}{2} \epsilon_{ij}
\psi_i \psi_j B  
\label  {eq: superwave}
\end{equation}
\noindent Where the four states  $\phi,A^i $ and B    are
functions of the coordinates $x_i$ only.On the  superspace the
hamiltonian is represented as.
\begin {equation} H=- \frac{ \partial^2}{\partial x_i\partial
x_i} +\frac{1}{2r^2} {( \frac{\partial V}{\partial \theta }
)}^2-\frac{1}{2} [ \frac{\partial}{\partial \psi_i},\psi_j ]
\frac{ \partial ^2 V}{ \partial x_i \partial x_j } \\
\label  {eq:hamiltonian1}
\end{equation}
\noindent The eigenvalue equation on the superspace 
\begin {equation} H \Phi=E\Phi
\label  {eq:schrodinger}
\end{equation}
\noindent is projected on each component and gives the set of
eigenvalue equations.
\begin {eqnarray}
		H B&=&E B \nonumber \\
	   H \Phi &=&E \Phi \nonumber \\
  ( H \delta_{ij} &+ &\frac{\partial^2 V}{\partial x_i \partial
x_j })A^i =E A^j    \nonumber \\
		   H&=&- \frac{ \partial^2}{\partial x_i \partial x^i}
+\frac{1}{2r^2} [ {( \frac{ \partial V}{  \partial \theta })}
^2-  \frac{\partial ^2 V }{\partial ^2 \theta } ]  
\label {eq:schrodingercompon}
\end{eqnarray}
\noindent All these equations are of the same type ( the third
one has to be diagonalized ) ,we therefore  restrict ourselves
to the ghost zero wavefunction $ \Phi \;$ for instance.Let us
look for a solution of the form.
\begin {equation} {\cal F}( r ) \Psi ( \theta )
\label  {eq: ansatz}
\end{equation}
\noindent Putting this into the $ \Phi $ equation in ~\ref
{eq:schrodingercompon} and separating the variables,we get.
\begin {eqnarray}
\ddot{\Psi }( \theta )  +(  \zeta ^2 - W  ) \Psi ( \theta )= 0 
\nonumber \\
\ddot{ \cal F } +\frac{1}{2}   \dot {\cal F} + ( \frac{
\zeta^2}{r^2} +2E ) {\cal F }=0  \nonumber  \\
 \;\; W={(\frac{\partial V}{\partial \theta })}^2-( \frac
{\partial ^2 V }{\partial ^2 \theta })  
\label {eq:diff}
\end{eqnarray}
\noindent Where $\zeta$ is a separation parameter.As topological
theories are based on the independence  on scales ( any ) ,the
second ( Bessel ) equation is relevant in this respect.We
therefore should elucidate some important aspects of the
solutions which has been  ignored or  simply underestimated in
reference \cite{kn:Baulieu}.The solutions
 to the Bessel equation in  ~\ref {eq:diff} are   .
\begin {equation} {\cal F} \propto J_\zeta ( \sqrt 2 E r )
\label {eq: Bessel}
\end {equation}
\noindent
$ J_\zeta$ is a Bessel function of order $\zeta$.One of he
properties of the Bessel function is  that it vanishes
asymptotically as the argument E goes to zero for  fixed r.This
means that the system has no  admissible ground state.The
hamiltonian being of the form $H \propto\{Q,\bar Q \} $, all
accessible  states $ ( E \neq 0 )$ are not $ Q,\bar Q $
invariant.The supersymmetry is therefore broken in a way that no
mass gap is introduced .This in turn opens the possibility of
having BRST-exact operators  with  non-vanishing expectation
values.,which in addition will be shown to be topological in the
sense that they do not depend on any scale present in the theory
.A second property which is very relevant to the definition of
topological observables is that Bessel functions are known to be
(ortho)normalizable only on a finite "volume".To define
orthonormality we introduce an infrared cut-off .This is a new
scale not present in the lagrangian.This length is defined such
that  ( We omit the  $ \sqrt 2  $ factor for convenience in what
follows) 
\begin {equation} J_\zeta ( EL )=0
\label  {eq: Bessel roots}
\end{equation}
\noindent For a fixed L this equation has an infinite set of
roots $ E_s$ .Bessel functions are thus  normalized consequently
as  \cite {kn:Smirnov}.
\begin {equation}
\int _0^L r J_\zeta ^2 ( E_s r ) dr= \frac{L^2}{2} J_{\zeta
+1}^2 ( E_s L )  
\label  {eq: orthogonality}
\end{equation}
\noindent This new scale will serve to properly define
topological invariants.
\section{ Calculability of observables and the Sturm-Liouville
system} Let us now concentrate on the angular dependance of the
solutions .This is given by the equation
\begin {eqnarray}
\ddot{\Psi }( \theta ) + ( \zeta ^2 - \alpha^2 -W ) \Psi = 0  
\nonumber  \\
 \;\;  W={(\frac{\partial \Phi}{\partial \theta })}^2-( \frac
{\partial ^2 \Phi }{\partial ^2 \theta }) 
\label {eq:angulardepen}
\end{eqnarray}
\noindent In the above equation the prepotential has been
replaced with the general solution  ~\ref{eq:prepotential} .We  
will distinguish two different cases of interest wich will be
shown to lead to different  topological informations.The first
case of vanishing potential is trivial and corresponds to the
potential $ W=0 $ 
\begin {equation}
\ddot{\Psi} ( \theta ) + ( \zeta ^2 -\alpha ^2) \Psi = 0  
\label  {eq: angular depen1}
\end{equation}
\noindent The solutions are simply $ \Psi=e^{in\theta} $  where
$ \zeta^2=n^2+\alpha^2  ,n\epsilon Z $ .The  second case  of
nonvanishing potenial is highly non  trivial.In this case the
prepotential is quite general and is only required to be
r-independant  and  periodic in $\theta$ as a consequence of the
cohomology of the target manifold. We hereafter , as far as the
solutions of the angular equation are concerned will understand
$\zeta^2 $ as shifted by $\alpha^2$ .Equation
~\ref{eq:angulardepen} with  W periodic is known as Hill's
equation \cite {kn:Hill}.We will show  in the next section that 
if one  makes  the judicious choice of a an even potential W
,only the existence and the orthonormality  of the solutions
will be needed for the computation of the vaccum  expectation
values of the  relevant topological invariants.Let us  convert 
Hill's equation to the well understood  Sturm-Liouville systems
in which  conditions for  the existence and the orthonormality
of the solutions are more  transparent .To this end we use the
periodicity property of the potential to write the  solutions as
Bloch wavefunctions.That is .
\begin {equation}
\Psi ( \theta ) =exp( i k\theta ) u_k ( \theta)
\;\;\;\;\;\;0\leq k <1
\label  {eq: Bloch}
\end{equation}
\noindent Where k 's are the eigenvalues of the translation
operator ( periodicity) and where $ u_k ( \theta ) 
$ are periodic functions with the same period as the potential
W.Putting this into ~\ref{eq:angulardepen} we get .
\begin {equation}
\ddot{u_k} +2ik \dot{u_k}+( \zeta^2-k^2-W ) u_k=0
\label  {eq: diff equation}
\end{equation}
\noindent As any second order ,linear ,homogeneous differential
equation,it can be transformed to the form.
\begin {eqnarray} L u_k&=&-\lambda  s u_k \nonumber \\
      L&=&\frac{d}{d\theta}( p( \theta) \frac{d}{d\theta} )+ q(
\theta )
\label  {eq: Sturm}
\end{eqnarray}
\noindent Where the $ p (\theta ) ,q (\theta ) $ functions and $
\lambda $ are defined as follows.
\begin {eqnarray}
 s( \theta )&=&e^{2ik\theta} \nonumber \\ p ( \theta )&=&e^{2ik
\theta} \nonumber \\ q ( \theta )&=&-W \; e^{2ik \theta} \\
\  lambda &=&-\;k^2 +\zeta^2  \nonumber 
\label  {eq: parameters}
\end{eqnarray}
\noindent This is a singular Sturm-Liouville system \cite
{kn:Kovach} with periodic boundary conditions .Any such system
with periodic boundary conditions is called regular and have
orthonormal solutions if the function $ p(\theta)$ is
continuously differentiable and 
\begin {eqnarray} p( \theta)& \geq& 0 \nonumber \\
    p ( 0 ) &   =  &p ( 2\pi ) 
\label  {eq: Sturm2}
\end{eqnarray}
\noindent The first condition fails  as  the function $ p
(\theta ) $  is oscillatory and thus may take positive as well
as negative values,the second ,because  k is real .We will
therefore limit ourselves to the set of orthonormal solutions
and put k=0 .This is the subset of periodic solutions. The
nonperiodic solutions  may anyhow be discarded on the basis that
they probably do not describe Hilbert spaces due to their  not 
being (orth)normal .In this subspace the Sturm-Liouville system
get regular and simplify to the form.
\begin {equation}
\ddot{u} + ( \zeta ^2 - W ) u = 0
\label  {eq:Sturm3}
\end{equation}
\noindent Note that this is of the same form as the starting
equation in  ~\ref{eq:angulardepen} but here the solutions are
now periodic.There is a set of theorems \cite {kn:Kovach} in the
chapter of boundary and eigenvalue  problems which state that
the above equation independently of the potential W provided it
is continuous ,has real eigenvalues and  all ordered as ( In
writing $\zeta^2$ in the equation above we have already
anticipated the positivity of the eigenvalues)
\begin {eqnarray} 0<\zeta^2_1<........<\zeta^2_p \nonumber \\
lim_{ p\rightarrow \infty }\;\;\zeta _p = \infty 
\label  {eq: eigenvalues}
\end{eqnarray}
\noindent And moreover  that the corresponding eigenfunctions
are orthonormal with weight 1 .As we  said already the exact
expressions for the eigenfunctions and values of the spectrum
will not be relevant to our purpose.It is clear that the the $
\Phi $ term in the  prepotential V can be chosen at will ( even
or odd ) provided it is periodic.If  $ \Phi $ is even then so is
the potential W   .$ \Phi $ and W have the same parity only if $
\Phi $ is even.On the other hand the operator $ \frac{
\partial^2}{\partial \theta^2}+ ( \zeta ^2 -W ) $ in
~\ref{eq:Sturm3} commutes with the parity operator    $ {\cal P}
u_\zeta ( \theta ) =u_\zeta ( -\theta ) $ .The solutions
$u_\zeta $  can then be chosen  as eigenfunctions of $ {\cal P}
$ as well,that is either even or odd functions of   $ \theta $ 
.Being periodic they could be expanded  on a general exponential
basis.If $ u_{\zeta p}$ belongs to the subset of even functions
for instance then it  can be expanded on the cosine basis only.
\begin {equation} u_{\zeta p}=\sum_{m=0}^\infty A_m^{\zeta _p
}\;\;cos \;m \theta 
\label  {eq: cosine expanssion}
\end{equation}
\noindent We are now ready to write down vaccum expectation
values for topological observables  .On dimensional grounds ,one
may select the following candidates ,together with their
hermitian conjugates.
\begin {eqnarray} {\cal O}_\theta &=&\{Q,\epsilon^{ij} x_i\bar
{\psi_j}\} =-i\partial _\theta +i \partial _\theta V \nonumber \\
       {\cal O}_r&=&\{ Q,ix_j\bar\psi_j \} = r \partial_ r  
\label  {eq: invariants}
\end{eqnarray}
\noindent In writing these expressions we dropped ghost terms as
they give vanishing actions on the $ \Phi $ wavefunctions we are
considering.Let us  first show that the second candidate is
discarded as it depends explicitly on energy.
\begin {eqnarray} {< {\cal O} _r >}_{E,\zeta}&=&\frac{ \int _0^L
J_\zeta \partial _r J_\zeta r^2 dr}{\int_0^L J_\zeta ^2 rdr}
\nonumber \\
			     &=&-1 + \frac{ J_\zeta ^2 ( E_s L ) }{J_{\zeta+1}^2 (
E_sL ) }
\label {eq:meanvalues}
\end{eqnarray}
\noindent This expectation value is indeed equal to -1 as we
have $ J_\zeta ( E_s L ) =0$ by construction .It seems then that
it depends neither on E nor on L,but this result is only true
for a denumerable set of energies $ E_s$, Any small deviation
from these energies would lead to different values for the
expectation value.To see this expand the right hand side of
~\ref{eq:meanvalues} around $ E_s$ .Put $ E=E_s+\epsilon $.
\begin {eqnarray} J_\zeta ( EL ) = J_\zeta ( E_sL ) + \epsilon
L  J'_\zeta ( E_sL ) \nonumber \\ = -\epsilon L J_{\zeta +1 } (
E_sL ) 
\label  {eq: invariants1}
\end{eqnarray}
\noindent Where we have used the relations $ J_\zeta ( E_sL )=0$
and $ J_\zeta ' (E_sL ) = - J_{\zeta +1} ( E_sL ) $ . $ J' $
being the derivative of J .Putting this into  ~\ref
{eq:meanvalues}  we get .
\begin {equation} {< {\cal O} _r >}_{E,\zeta} \approx -1 +
\epsilon^2 L^2
\label {eq: invariants2}
\end{equation}
\noindent Which shows an explicit dependance on energy and on
L.The other candidate turns out to be a topological
invariant.The expectation value of $\cal O$  ( we drop the index
) between orthonormal states $u_{\zeta p}$ is.
\begin {equation} {< {\cal O } >}_{E,\zeta} =\frac{ \int
u^*_\zeta \;\;( -i \partial_\theta +i\partial _\theta V )
\;\;u_\zeta d\theta }{\int u^*_\zeta u_\zeta d\theta }
\label {eq:invariants3}
\end{equation}
\noindent In the case of a vanishing potential for which the
solutions are  $ u_n=e^{ in\theta } $  we  get 

\begin {equation} {< {\cal O}>}_{E,n} =n+i\alpha
\label  {eq:result1}
\end{equation}
\noindent In the case of non vanishing potentials the
orthogonality of the basis cancels the $  \partial_\theta  $
contribution as we have.
\begin {equation}
\int u^*_{\zeta } \partial_\theta u_{\zeta } d\theta =0
\label  {eq: orthogonality1}
\end{equation}
\noindent Only the  c-number term in ~\ref {eq:invariants3} 
will  then contribute,hence .
\begin {eqnarray} {<{\cal O} >}_{E,\zeta}&=&i\partial_ \theta V
\nonumber \\
				&=&-\sum_{m\epsilon Z/( 0) } \frac{ \rho ( m ) }{m} e
^{im\theta } +i\alpha 1
\label  {eq:result2}
\end{eqnarray}
\noindent Where the "spectral" function $ \rho ( m ) $ is even
in m as V is even in $ \theta $ . The potential being periodic
we have  resolved  it into its  Fourier components.It is an
imortant fact that V may be selected with even parity as this
forced the solutions to be of a given parity and thus to discard
the unwanted contribution coming from the derivative operator (
angular momenta).In the next section we will give a further
justification  of  the selected parity for the prepotential  as
a necessary requirement for proper  identification of the
observables as topological invariants of the punctured plane.

\section{ Twisting of the homotopy structure and topological
interpretation} Now to  interpret the Q-exact invariants in
terms of the invariants which live naturally on the punctured
plane we should first recall some known results and modify the
homotopy structure somehow.The topology of the punctured plane
is encoded within the fundamental group of homotopy $ \Pi_1$
which is isomorphic to the set of integers,$ \Pi_1 \cong Z  $  .
Thus each topological state is labelled by the integers and will
be denoted $ \mid n >$ .We then have a set  of operators at our
disposal W the winding number and $ \Pi_1 ( m ) $ the elements
of the homotopy group which act on them .The whole topological
information may be summarized in the following set of equations.
\begin {eqnarray}
				     W \mid n >&=&n \mid n > \nonumber \\
			  Pi_1 ( m ) \mid n >&=& \mid n+m > \\
\sum _{n\epsilon Z } \mid n ><n \mid &=&1 \nonumber 
\label {eq:homotopy}
\end{eqnarray}
\noindent The group $ \Pi_1$ being abelian , all its elements
are generated by $ \Pi_1 ( 0 )=1 $ the identity and the
generator $ \Pi_1 ( 1 ) $ for instance.
\begin {equation}
\Pi_1 ( m )=\Pi_1 ^m ( 1 ) ,\;\;\;\;m\epsilon Z
\label  {eq: homotopy generator}
\end{equation}
\noindent To make the connection with Q-exact invariants we
convert to the theta basis.The fundamental group being abelian
,its unitary irreducible representations are one
dimensional.Write these states as $ \mid  \theta > $ where $
\theta $ is an angle.The operators $ \Pi^s_1 $ and W act on
these states as.
\begin {eqnarray}
\Pi _1( m )\mid \theta >&=&e^{im\theta} \mid \theta >\nonumber \\
	  W \mid \theta > &=&i \partial _\theta \mid \theta > 
\label  {eq:irreductible states}
\end{eqnarray}
\noindent On the other hand both kinds of states are related to
each other through the Fourier transform.
\begin {eqnarray}
\mid \theta >&=&\sum_{m\epsilon Z} e^{-in\theta} \mid n > 
\nonumber \\
\mid n >&=&\int _0^{2\pi} e^{in\theta} \mid \theta >
\frac{d\theta}{2\pi} 
\label  {eq:fourier transf}
\end{eqnarray} At this stage one could not propose to identify
the resulting topological invariants in terms of the homotopy of
the target manifold.That is to identify in the case at hand  the
invariants in terms of the winding number W and the elements of
the  fundamental group of homotopy $\Pi_1 ^ s( m ) $ .To make
the identication properly one should perform a twisting of the
homotopy structure.In an unpublished work \cite {kn:Mekhfi} we 
performed such twisting .What we mean by that is that by viewing
momentarily topological loops  as intrinsec physical objects we
allow them to have self interactions .The reason one has to do 
that is that the invariants to be identified  being Q-exact
depend explicitly on  the coupling  constants present in the
theory as $ Q \propto \partial V$  but these extra parameters (
coupling  constants ) are not ( present ) needed to characterize
the (algebraic) topology of the punctured  plane.One should
enrich the homotpy structure somehow to implement these
parameters.We will  just quote the results which could be
checked without much effort.Starting from the operators W  and $
\Pi ( m ) $  (Hereafter we omit the subscript 1 ),we may build a
new  functional operator $  W_\lambda[\rho ]$  an " effective
winding number "as .
\begin {eqnarray} W_\lambda [\rho] &=& W+\lambda \sum_{m\epsilon
Z }\rho ( m ) \Pi ( m ) \nonumber \\
			  &=&\;\; e^{\lambda J}\;W \;e^{- \lambda J}  +\lambda \rho (
0 ) \\
	    With  \;\;\;    J[\rho]\;&=&-\;\sum_{m\epsilon Z/( 0 )} 
\rho( m ) \frac{\Pi(m)}{m} \nonumber
\label  {eq:twisting}
\end{eqnarray}
\noindent To get the first equation from the second one ,one may
use the commutation relation $ [ W,\Pi  (m) ]=m\Pi (m)$ one may
extract from the set of defining equation ~\ref{eq:homotopy}.Let
us first  make clear the choice of the special interaction in
\ref {eq:twisting} by looking at the second equation .The first 
term ( set   $ ( \rho ( 0 ) =0$ ) is just the twisting of the
winding number operator W.This is  the analog of the twisting of
the exterior derivative operator first performed by Witten
\cite  {kn:Witten1} in order  to introduce the prepotential in a
cohomological way let us say.In the above formula we did the 
same thing but in a homological way.  ( Recall that the
isomorphy $ H_1 \cong \Pi_1 $  means  that homology and homotopy
are equivalent words in the present case ).So one has to compare
the above twisting with 
\begin {eqnarray} d_\lambda[V]=e^{\lambda V } d \;\; e^{-\lambda
V}
\label  {eq: twisting1}
\end{eqnarray}
\noindent Where $ d \equiv W = i \partial_\theta $ ( on the  $
\mid \theta >$ states ).With this we see that we
 have done more than a twisting as we added an affine 
non-vanishing term $ \lambda \rho ( 0 ) 
$.This term is  fundamental as it permits to define not only a
twisted version  of the system but  another physically different
one.It is interesting to regard the operator $ W_\lambda [\rho]
$ as a
 hamiltonian which describe interacting homotopic loops with. W
as  the unperturbed part and  where  $\lambda \rho ( m ) $ is an
m-dependant coupling constant function.In the above cited 
reference  \cite {kn:Mekhfi}  we solved the eigenvalue problem
associated to the above  hamiltonian and find that it possesses
a set of eigenstates  all well defined and having real 
eigenvalues  to which only the affine term contributes.$
n_{\lambda \rho ( 0 ) }= n+\lambda \rho ( 0 ) $  with $ 
n\epsilon Z $ and $ 0 <\lambda \rho <1 $ .In the case of pure
twisting $ \rho ( 0 )=0 $  the  spectrum remains unchanged as 
we know that the twisting operation is a kind of symmetry for 
the system.The  new eigenstates are related to the old ones
through the formula .
\begin {equation}
\mid n,\lambda ,\rho > = e^{ \lambda J} \mid n >
\label  {eq:lambdastates}
\end{equation}
\noindent These states  describe interacting homotopic loops and
may be interpreted as  a new larger basis for the homotopy 
structure which reduces to the usual homotopy basis once we set
$ \lambda =0 $ .But to get a topologically interesting system we
further require that the hamiltonian $ W_\lambda [\rho] $  be
hermitian with the function $ \rho ( m ) $  being real.This 
fixes the latter function to be an even function in m that is $
\rho ( -m )=\rho ( m ) $.As a consequence the operator $
e^{\lambda J }$ is  unitary ( J antihermitian ) .Has this 
operator  being  not unitary which  could arise if $ \rho$ is
odd or having no parity the new states which will no longer be
the unitary transform of the old ones .This would certainly
spread out the  initial topological information about the
homotopy structure available before we switch  on the 
interaction.This observation should be seen as a further ( here 
topological ) justification of why  we adopted periodic
potentials of even parities which was translated on the $ \rho $
function in  ~\ref{eq:result2}  being an even function of
m.Recall  that this condition was necessary and  sufficent to
get rid of the unwanted  contributions coming from the angular
momenta  which would require full knowledge of the spectrum and
the associated eigenfunctions. We may now give the topological
interpretation .For the two  cases  the  identifications are as
follows. In the  first case  we get .
\begin {equation}
	    {\cal O}\; = \;W+i\alpha \;1 
\label  {eq: identification1}
\end{equation}
\noindent This is  the effective winding number $ W_\lambda
[\rho] $ in ~\ref {eq:twisting} with the opertor J set to zero
and  where $ \lambda \rho ( 0 ) =\alpha $ We see in this case
that there is no need to twist the winding  number .For the 
second case  for which the prepotential is  periodic ,even and
the nonperiodic  solutions discarded  on comparing  the
expectation value of the Q -exact invariant in equation ~\ref
{eq:result2}  with that of the J operator taken between $ \mid
\theta > $ states the  identification is.
\begin {equation}
	    {\cal O}\; =\;J + i\alpha \;1
\label  {eq: identification2}
\end{equation}
\noindent To end up we may quote another result from our recent
work \cite {kn:Mekhfi1}on the topological  setting of Bessel
functions  in order to give a further meaning to the invariant
J.We have shown  there that if one associate a reduced Bessel
function $ j_n( z )=\frac{J_n( z ) }{ z^n} $ to the loop  state
$\mid  n > $ then picking the particular value for the coupling
constant $ \rho ( m ) = ( - )^m 
 $ one may rewrite the formula ~\ref {eq:lambdastates} as .
\begin {equation} j_{n+\lambda}=\;\;e^{\lambda \;J }\;\; j_n
\label  {eq:generator Bessel}
\end{equation}
\noindent Where we can see that the invariant J we got  ,for the
typical choice of the spectral function $\rho $ is the generator
of real order reduced Bessel  functions  ( from integer order
ones ).

\end{document}